\documentstyle[sprocl,epsf]{article}

\def\be{\begin{equation}}
\def\ee{\end{equation}}
\def\bea{\begin{eqnarray}}
\def\eea{\end{eqnarray}}

\newcommand{\bef}{\begin{figure}}
\newcommand{\eef}{\end{figure}}
\newcommand{\hmp}{ h^{-1}Mpc}
\newcommand{\etal}{{\it et al.}}
\def\spose#1{\hbox to 0pt{#1\hss}}
\def\ltapprox{\mathrel{\spose{\lower 3pt\hbox{$\mathchar"218$}}
 \raise 2.0pt\hbox{$\mathchar"13C$}}}
\def\gtapprox{\mathrel{\spose{\lower 3pt\hbox{$\mathchar"218$}}
 \raise 2.0pt\hbox{$\mathchar"13E$}}}
\def\inapprox{\mathrel{\spose{\lower 3pt\hbox{$\mathchar"218$}}
 \raise 2.0pt\hbox{$\mathchar"232$}}}

\begin{document}

\title{The Visible Universe at the Light of Modern Statistical Physics}

\author{L. Pietronero,F. Sylos Labini and  M. Montuori}

\address{Dipartimento di Fisica, Universit\`a di Roma
``La Sapienza''\\
P.le A. Moro 2, I-00185 Roma, Italy.}

\maketitle
\abstracts{
In the last years there has been a growing interest in the 
understanding a vast variety of
scale invariant and critical phenomena occurring in nature. 
Experiments and observations
indeed suggest that many physical systems develop spontaneously 
correlations with
power law behaviour both in space and time. Pattern formation, 
aggregation phenomena,
biological systems, geological systems, disordered materials, 
clustering of matter in the
universe are just some of the fields in which scale invariance has 
been observed as a common and basic feature.
However, the fact that certain structures exhibit fractal and complex 
properties does not tell us why this happens. 
A crucial point to understand is therefore the origin 
of the general scale-invariance of natural phenomena. 
This would correspond to the understanding
of the origin of fractal structures and of the properties of Self-
Organized Criticality (SOC) from the knowledge of the microscopic 
physical processes at the basis of these phenomena.
Fractal 
geometry can play a
crucial role in extracting the correct physical properties from 
experimental data. In particular, 
the recent availability of complete three dimensional samples
of galaxies and clusters permits a direct  study of their 
spatial properties. We present a brief review  of  
galaxy correlations based on the methods of modern Statistical Physics.
These methods are able to identify  self-similar and non-analytical properties, and 
allow us to test the usual homogeneity 
assumption of luminous matter distribution.
 The new analysis shows that 
all the available data are consistent with each other and 
show fractal correlations (with dimension $D \simeq 2$) 
up to the deepest scales probed until now 
($1000 \hmp$) and even more as indicated from the new interpretation 
of the number counts.
}
  
\section{Introduction}

In scale invariant phenomena, events and information spread over 
a wide range of length
and time scales, so that no matter what is the size of the scale 
considered one always
observes surprisingly rich structures. These systems, with very 
many degrees of
freedom, are usually so complex that their large scale behaviour 
cannot be predicted from
the microscopic dynamics. New types of collective behaviour arise 
and their understanding represents one of the most challenging areas in 
modern statistical physics.

Nowadays the physics of scale invariant and
complex systems is a novel field which is including topics from 
several disciplines
ranging from condensed matter physics to geology, biology, 
astrophysics and
economics. This widespread interdisciplinary corresponds 
to the fact that these new ideas allow us to look at natural phenomena 
in a radically new and original way, eventually leading to unifying concepts 
independently of the detailed structures  of systems.
The interest in the field of scale free phenomena and complex systems 
has been largely due to two factors. First the emerging
availability of high powered computers over the past decade has 
enabled to readily
simulate complex and disordered systems. Second the cross 
disciplinary mathematical
language for describing these phenomena evolving under 
conditions far from equilibrium
has only become available in the past years.
The study of critical phenomena in second order transitions 
introduced the concepts of
scaling and power law behavior. Fractal geometry provided the 
mathematical framework
for the extension of these concepts to a vast variety of natural 
phenomena. The physics of complex systems, however, turned out to be 
effectively new with respect
to critical phenomena. The theory of equilibrium statistical physics 
is strongly based on
the ergodic hypothesis and scale invariance develops at the critical 
equilibrium between
order and disorder. Reaching this equilibrium requires the fine 
tuning of various
parameters. On the contrary most of the scale free phenomena 
observed in nature are
self-organized, in the sense that they spontaneously develop from 
the generating
dynamical process. One is then forced to seek the origin of the scale 
invariance in nature
in the rich domain of nonequilibrium systems and this requires the 
development of new
ideas and methods.

In order to identify phenomenologically which microscopic 
dynamics may lead to fractal
and SOC properties, it is necessary to define simple physical models 
that should capture
the essential ingredients of these phenomena. In the past years a 
large number of models
have been introduced and studied, mostly by numerical 
simulations. 
Given this scenario the aim of the current studies on fractal and self-organized 
systems   is twofold.
\begin{itemize}

\item Although fractal growth models and SOC have
provided a useful insight into a vast array of problems, many 
important questions are
still open. In fact it is particularly important to reach a more complete and 
predictive theoretical
understanding of fractal growth and the SOC mechanism. These 
issues are still
unresolved, the present picture being based on the analysis of 
particular models. In this
sense, it is still missing a general and precise definition of the 
circumstances leading to
fractals and  SOC and the identification of common features in 
different systems.
For instance  a crucial issue is the role of universality in fractal and 
SOC phenomena. In
usual critical phenomena, the same exponents that define the onset 
of magnetisation, also
describe the liquid vapor transition in water. This strong 
universality appears to be a
characteristic of equilibrium systems. Self-organized systems, on 
the other hand, do not
seem to exhibit the same degree of universality as the fractal 
dimension can be easily
altered by simple changes in the growth process. This lack of 
universality is sometimes
viewed as a negative element because one is forced to describe 
specific systems instead
 of a single universal model. The truth is probably the opposite. Some theoretical
concepts can be considered as general or universal, but the 
inherent diversity of the
various models that have been studied, adds another fascinating 
dimension in the intellectual search. 
This is a
difficult and deep question because, in addition to its intrinsic 
interest,  it has deep
implications on the validity of the simplified computer models with 
respect to physical
reality. Clearly such a problem can only be investigated by a 
comprehensive effort that
involves computer simulations, analytical tools and suitably 
designed experiments. 

\item
While the theoretical activity is focused mainly on simple 
cellular automata or toy models, it is strikingly important to 
understand the relations between theory and real experiments.
Therefore, inspired from the early generation of prototype 
models one intends to
formulate more realistic models for a detailed interpretation of 
specific phenomena in
various fields. These models should be concrete tools 
for the understanding of the phenomenology and the
characterization of several experimental problems.
On the other hand, there is a need of real experiments which 
can discriminate among the various theoretical framework.
One  also expects that experiments could point out new 
properties and new models
that may enlighten specific theoretical questions.
\end{itemize}

 The 
introduction of new
ideas, inspired by fractal geometry and scaling, irreversible and 
non ergodic dynamical
systems leading to self-organization and stochastic processes of 
various types, give rises to a
considerable enrichment of the traditional framework of Statistical physics and provides 
efficient methods for
characterizing and understanding complex systems.

The impact of fractals in physics can be assessed along three 
different lines of increasing
complexity:
\begin{itemize}
\item[{\bf (a)}]
Fractal geometry merely as a mathematical framework which leads 
to a re-analysis of
 known data that results in a revamping of long standing points of 
view.
\item[{\bf (b)}] The development of physical models for systems that 
exhibit fractal and SOC behaviour.
\item[{\bf (c)}] The construction of physical theories that allow us to 
understand the origin of fractal
structures and SOC properties in various systems.
An additional question which  refers to all the previous points, is the 
study of the physical
properties of fractal and SOC systems such as transport, vibrational, 
electronic properties etc.
\end{itemize}

The first physical model of fractal growth was the Diffusion Limited Aggregation (DLA)
introduced in 1981 by Witten and Sander (see for a review \cite{evp95}). 
This model was 
generalized by Niemeyer,
Pietronero and Wiesmann \cite{npw}with the Dielectric Breakdown Model 
(DBM), inspired by
discharges in gases, that also clarified the underlying mathematical 
properties based on
Laplace equation. In this respect, it may be surprising to see that 
the Laplace operator,
which usually leads to smooth properties, in the case of these 
models drives
spontaneously its boundaries into strongly irregular fractal 
structures. These Laplacian
fractals are believed to capture the essential properties of a variety 
of phenomena such as
electrochemical deposition, dielectric and 
mechanical breakdown, viscous fingering in fluids,  
propagation of fractures
 and various properties of colloids.

More recently, in order to put in a broader framework the self-
organization properties of
the above models, Bak, Tang and Wiesenfeld \cite{btw} invoked the concept 
of Self-Organized
Criticality (SOC) as a unifying framework to describe a vast class of 
dynamically driven 
systems which evolve spontaneously in a stationary state with a 
broad power law
distribution of discrete energy dissipating events. 
In these models criticality seems to emerge automatically 
without the fine
tuning of parameters. Because of the enormous conceptual power, 
SOC ideas have
invaded rapidly throughout the sciences, from physics and 
geophysics to biology and
economics, as a prototype mechanism to understand the 
manifestation of scale invariance
and complexity in natural phenomena.

An important area in which fractal 
geometry can play a
crucial role consists in extracting the correct physical properties from 
experimental data of the
galaxy distribution in space. The usual analysis measures the 
deviations of the
conditional density at a given distance from the average density. 
Note however that this
procedure implicitly assumes homogeneity and thus cannot 
objectively test if the
considered portion of the universe is or not homogeneous. Some 
years age  we 
proposed a more general method of analysis based on the concepts 
nd methods of
modern statistical physics \cite{cp92} \cite{slmp97}. 
This led to the surprising result that 
galaxy correlations are
fractal and not homogeneous up to the limit of the available data. 
All the structures
observed in the universe appear to be characterized by strongly 
irregular, scale invariant
fluctuations. These results led to a large debate in the field and the 
new data, expected to
appear in the near future, should provide a definitive test of these 
properties. These
results may lead to fascinating conceptual implications about our knowledge of the
universe and to a new scenario about the theoretical challenge in 
this field.

\section{Self-similarity and power law correlations}

A fractal consists of a system in which more
and more structures appear at smaller and
smaller scales and the structures at small
scales are similar to the one at large scales.
 Starting from a
 point occupied by an object we count how 
many objects are present within a volume 
characterized by a certain length scale in
 order to establish a generalized "mass-length" 
relation from which one can define the fractal 
dimension.
   We can then write a relation
 between $\:N$ ("mass") and $\:r$ ("length") of type \cite{man83}:
\be
\label{l2}
N(r) = B\cdot r^{D}
\ee
where the fractal dimension is $D$ and the prefactor $\:B$ 
is instead related to the lower cut-offs.
It should be noted that Eq.\ref{l2} corresponds 
to a smooth convolution of a strongly 
 fluctuating function. Therefore a fractal 
 structure is always connected with large
 fluctuations and clustering at all scales.
 From Eq.\ref{l2} we can readily compute the 
 average density $\:<n>$ for a spherical sample of
 radius $\:R_{s}$ which contains a portion 
 of the fractal structure: 
\be
\label{l5}
<n> =\frac{ N(R_{s})}{V(R_{s})} = \frac{3}{4\pi } B R_{s}^{-(3-D)}
\ee
From Eq.\ref{l5} we see that the average density 
is not a meaningful concept in a fractal 
because it depends explicitly on the sample 
size $\:R_{s}$. We can also see that for 
$\:R_{s} \rightarrow \infty$ the average density 
$\:<n> \rightarrow  0 $, 
therefore a fractal structure is asymptotically  
dominated by voids.

It is useful to introduce the conditional density from an point 
occupied as:
\be
\label{l6}
\Gamma (r)= S^{-1}\frac{ dN(r)}{dr} = \frac{D}{4\pi } B r ^{-(3-D)}
\ee
where $\:S(r)$ is the area of  a spherical shell of radius $\:r$.
Usually the exponent that defines the decay
 of the conditional density $\:(3-D)$ is called 
the codimension and it corresponds to the 
exponent $\:\gamma$ of the galaxy distribution.
 
We can now describe  how to perform the correct correlation
analysis which  can be applied in the case 
of an irregular distribution as well as of a regular one. 
We may start recalling  the 
concept of correlation. If the presence of an object at the point $r_1$ 
influences the probability of finding another object 
at $r_2$, 
these two points are correlated. Therefore there is a correlation
at  $r$ if, on average
\be
\label{e324}
G(r) = \langle n(0)n(r)\rangle   \ne \langle n\rangle  ^2.
\ee
where we average over all occupied points chosen as origin.
On the other hand there is no correlation if
\be
\label{e325}
G(r) \approx \langle n\rangle  ^2.
\ee
The physically meaningful definition of  $\lambda_0$ 
is therefore the length scale which separates correlated regimes from
uncorrelated ones.

In practice, it  is useful 
to normalize the correlation function (CF) 
of Eq.\ref{e324}
to the size  of the
sample under analysis. Then we use, following Coleman \& Pietronero \cite{cp92}
\be
\label{e326}
\Gamma(r) = \frac{<n(r)n(0)>}{<n>} = \frac{G(r)}{<n>}
\ee
where $\:<n>$ is the average density of the sample.  We stress
that this normalization does not introduce any bias even if the average
density is sample-depth dependent,
as in the case of fractal distributions,
because it represents
only an overall normalizing factor. 
In order to compare results from different catalogs
it is however more useful to use $\Gamma(r)$, in which
the size of a catalog only appears via the combination
$N^{-1}\sum_{i=1}^{N}$ so that a larger sample 
volume only enlarges the statistical sample over which averages are taken.
 $G(r)$ instead 
has an amplitude that is an explicit function of the sample's size
scale.

The CF of Eq.\ref{e326}
can be computed by the following expression
\be
\label{e327}
\Gamma(r) = \frac{1}{N} \sum_{i=1}^{N} \frac{1}{4 \pi r^2 \Delta r}
\int_{r}^{r+\Delta r} n(\vec{r}_i+\vec{r'})d\vec{r'} = 
\frac{BD}{4 \pi} r^{D-3}
\ee
where the last equality follows from Eq.\ref{l6}.
This function measures the average density at distance $\:\vec{r}$ from an
occupied point at $\vec{r_i}$
and it is called the {\it conditional density} \cite{cp92}.
If the distribution is fractal up to a certain distance $\lambda_0$,
and then it becomes homogeneous, we have that
$$
\Gamma(r) = \frac{BD}{4 \pi} r^{D-3} \;  r < \lambda_0
$$
\be
\label{e327b} 
\Gamma(r)= \frac{BD}{4 \pi} \lambda_0^{D-3} \; r \geq \lambda_0
\ee

It is also very useful to use the conditional average density in the 3-d space 
\be
\label{e328}
\Gamma^*(r) = \frac{3}{4 \pi r^3} \int_{0}^{r} 4 \pi r'^2 \Gamma(r') dr' =
\frac{3B}{4 \pi} r^{D-3} \; .
\ee
This function would produce an artificial smoothing of
rapidly varying fluctuations, but it correctly
reproduces global properties   \cite{cp92}.

For a fractal structure, $\Gamma(r)$ has a power law behaviour
and the conditional average density $\Gamma^*(r)$ has the form
\be
\label{e329}
\Gamma^*(r)= \frac{3}{D} \Gamma(r) \; .
\ee
For an homogenous distribution ($D=3$) these two functions
are exactly the same and equal to the average density.   

 Pietronero  and collaborators \cite{pie87} \cite{cps88} 
\cite{cp92} 
have clarified some crucial points of the
standard correlations analysis, and in particular they have discussed the 
physical meaning
of the so-called {\it "correlation length"}
  $\:r_{0}$ found with the standard
approach \cite{pee80} \cite{dp83} and defined by the relation:
\be
\label{e330}
\xi(r_{0})= 1
\ee
where
\be
\label{e331}
\xi(r) = \frac{<n(\vec{r_{0}})n(\vec{r_{0}}+ \vec{r})>}{<n>^{2}}-1
\ee
is the two point correlation function used in the standard analysis.
The basic point that \cite{cp92} stressed,
is that the mean density, $<n>$,
used in the normalization of $\:\xi(r)$, is not a well defined quantity
in the case
of self-similar distribution and it is a direct function of the sample size.
Hence only in the case that
the homogeneity  has been reached well within the sample
limits the $\:\xi(r)$-analysis is meaningful, otherwise
the a priori assumption of homogeneity is incorrect and the
characteristic lengths, like $\:r_{0}$, became spurious.

 For example 
 the expression of the $\:\xi(r)$ in the case of
 fractal distributions \cite{cp92} is:
\be
\label{e332}
\xi(r) = ((3-\gamma)/3)(r/R_{s})^{-\gamma} -1
\ee
where $\:R_{s}$ is the depth of the spherical volume where one computes the
average density from Eq.\ref{l5}.
From Eq.\ref{e332} it follows that

i.) the so-called correlation
length $\:r_{0}$ (defined as $\:\xi(r_{0}) = 1$)
is a linear function of the sample size $\:R_{s}$
\be
\label{e333}
r_{0} = ((3-\gamma)/6)^{\frac{1}{\gamma}}R_{s}
\ee
and hence it is a spurious quantity without  physical meaning but it is
simply related to the sample finite size.

ii.) $\:\xi(r)$ is power law only for
\be
\label{e334}
((3-\gamma)/3)(r/R_{s})^{-\gamma}  >> 1
\ee
hence for $r \ll r_0$: for larger distances there is a clear deviation
from the power law behaviour due to the definition of $\xi(r)$.
This deviation, however, is just due to the size of
 the observational sample and does not correspond to any real change
of the correlation properties. It is clear that if one estimates the
 exponent of $\xi(r)$ at distances $r \ltapprox r_0$, one
 systematically obtains a higher value of the correlation exponent
 due to the break of $\xi(r)$ in the log-log plot.

The analysis
 performed by $\xi(r)$ is therefore mathematically inconsistent, if
 a clear cut-off towards homogeneity has not been reached, because
 it gives an information that is not related to the real physical
 features of the distribution in the sample, but to the size of the
sample itself.

\section{Correlation properties of galaxy distribution} 

The main data of our correlation analysis 
are collected in Fig.\ref{fig1} (left part)
 in which we report the 
{\it conditional density as a function of scale}
 for the various catalogues. 
The relative position of the various behaviours of the conditional
density in different catalogs,  is not arbitrary but it is fixed 
by the luminosity function (a part for the cases of 
IRAS and SSRS1 for which this is 
not possible). The properties derived from different 
catalogues are compatible with each other and show a {\it power law 
decay} for the conditional density from $1 \hmp$ to $150 \hmp$
 without any tendency towards homogenization (flattening). This 
implies necessarily that the value of $r_0$ (derived from the $\xi(r)$ 
approach) will scale with the sample size $R_s$ as shown also from the 
specific data about $r_0$ of the various catalogues. The 
behaviour 
observed  corresponds to a fractal structure with dimension $D \simeq 
2$. The smaller value of CfA1 was due to its limited size. An 
homogeneous distribution would correspond to a flattening of the 
conditional density which is never observed
\begin{figure}
\epsfxsize 8cm
\centerline{\epsfbox{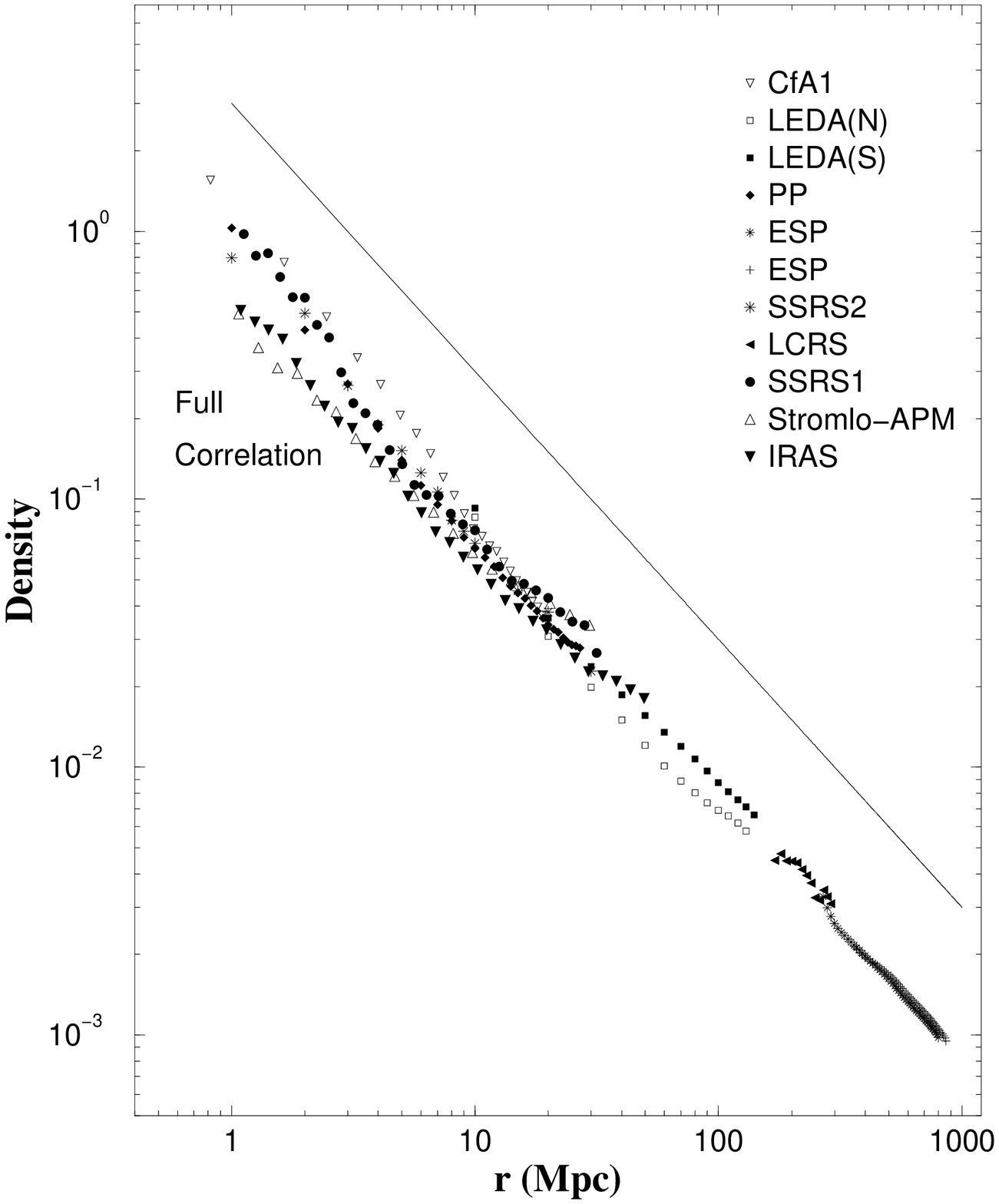}}
\caption{\label{fig1} Full correlation analysis for the various
 available redshift surveys in the range of distance $0.5 \div 1000
 \hmp$. A reference line with slope $-1$ is also shown, that 
corresponds to fractal dimension $D = 2$. 
 } 
\eef 
It is
 remarkable to stress that the amplitudes and the slopes of the
 different surveys match quite well. From this figure we conclude
 that galaxy correlations show very well defined fractal properties
 in the entire range $0.5 \div 1000 \hmp$ with dimension $D = 2 \pm
 0.2$. Moreover all the surveys are in agreement with each other.

 It is interesting to compare the analysis of Fig.\ref{fig1} with 
the usual one, made with the function $\xi(r)$, for the same 
galaxy catalogs. This is reported in Fig.\ref{fig2}
\bef
 \epsfxsize 8cm 
\centerline{\epsfbox{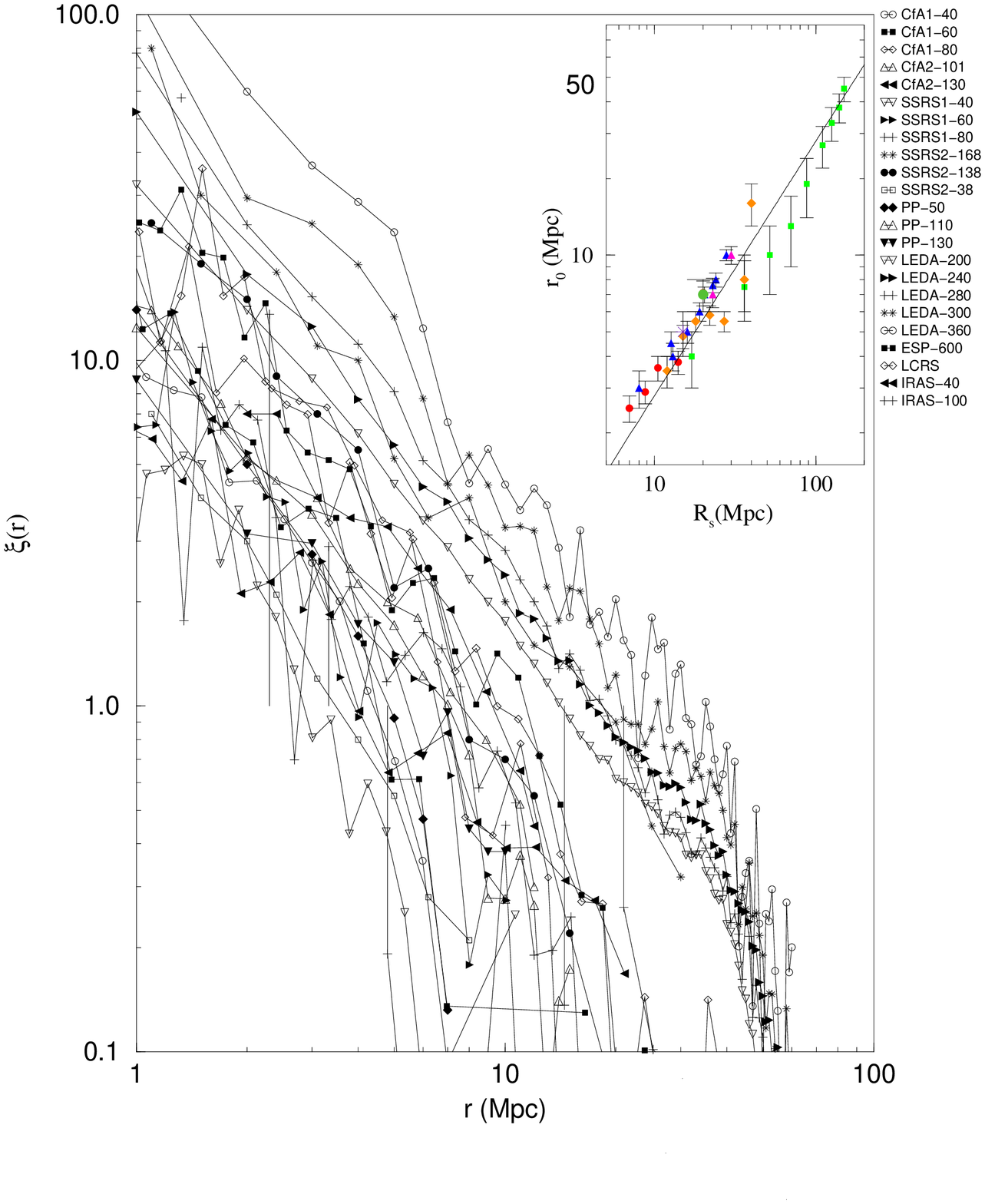}}  
\caption{\label{fig2}
Traditional analyses based on the function $\xi(r)$
of the same galaxy catalogs of the previous figure.
 The usual 
analysis is based on the a priori untested assumptions of 
analyticity and homogeneity. These properties
are not present in the real galaxy distribution and 
the results appear therefore rather confusing. 
This lead to the impression that galaxy catalogs are not good
enough and to a variety of theoretical problems like the 
galaxy-cluster mismatch, luminosity segregation, linear and 
non-linear evolution, etc.. This situation changes completely and 
becomes quite clear if one adopts the more 
general conceptual framework that is at the basis 
the previous figure}
\eef 
and, from this point of view, 
the various data appear to be in strong disagreement with 
each other. This is due to the fact that the usual analysis
looks at the data from the prospective of analyticity and large
scale homogeneity (within each sample). These properties have never
been tested, and they are not present in the real galaxy
distribution, so the result is rather confusing (Fig.\ref{fig2}).
Once the same data are analyzed with a broader perspective, the
situation becomes clear (Fig.\ref{fig1}) and the data of 
different catalogs result in agreement with each other. It is 
important to remark that analyses like those of Fig.\ref{fig2}
have had a deep influence in the field in various ways: 
first, in the standard analysis, 
the different catalogues appear in conflict with each other.
This has generated the concept of {\it not fair samples} and a 
strong mutual criticism about the validity of the data 
between different authors. In the other cases the
discrepances observed in Fig.\ref{fig2} have been 
considered real physical problems for which various technical
approaches have been proposed. These problems are, for example, 
the galaxy-cluster mismatch, luminosity segregation, 
the richness-clustering relation and 
 the linear non-linear evolution of the perturbations 
corresponding to the {\it "small"} or  {\it "large"}
amplitudes of fluctuations. We can now see that all this problematic
situation is not real and it arises only from a 
statistical analysis based on inappropriate and too restrictive
 assumptions that do not find any correspondence in the 
physical reality. It is also important to note that,
even if the galaxy distribution would eventually became
homogeneous at larger scales, the use of the above statistical
concepts  is anyhow inappropriate for the range of scales 
in which the system shows fractal correlations as those 
shown in Fig.\ref{fig1}.

\section{Conclusions and theoretical implications}

\bef
\epsfxsize 9cm 
\centerline{\epsfbox{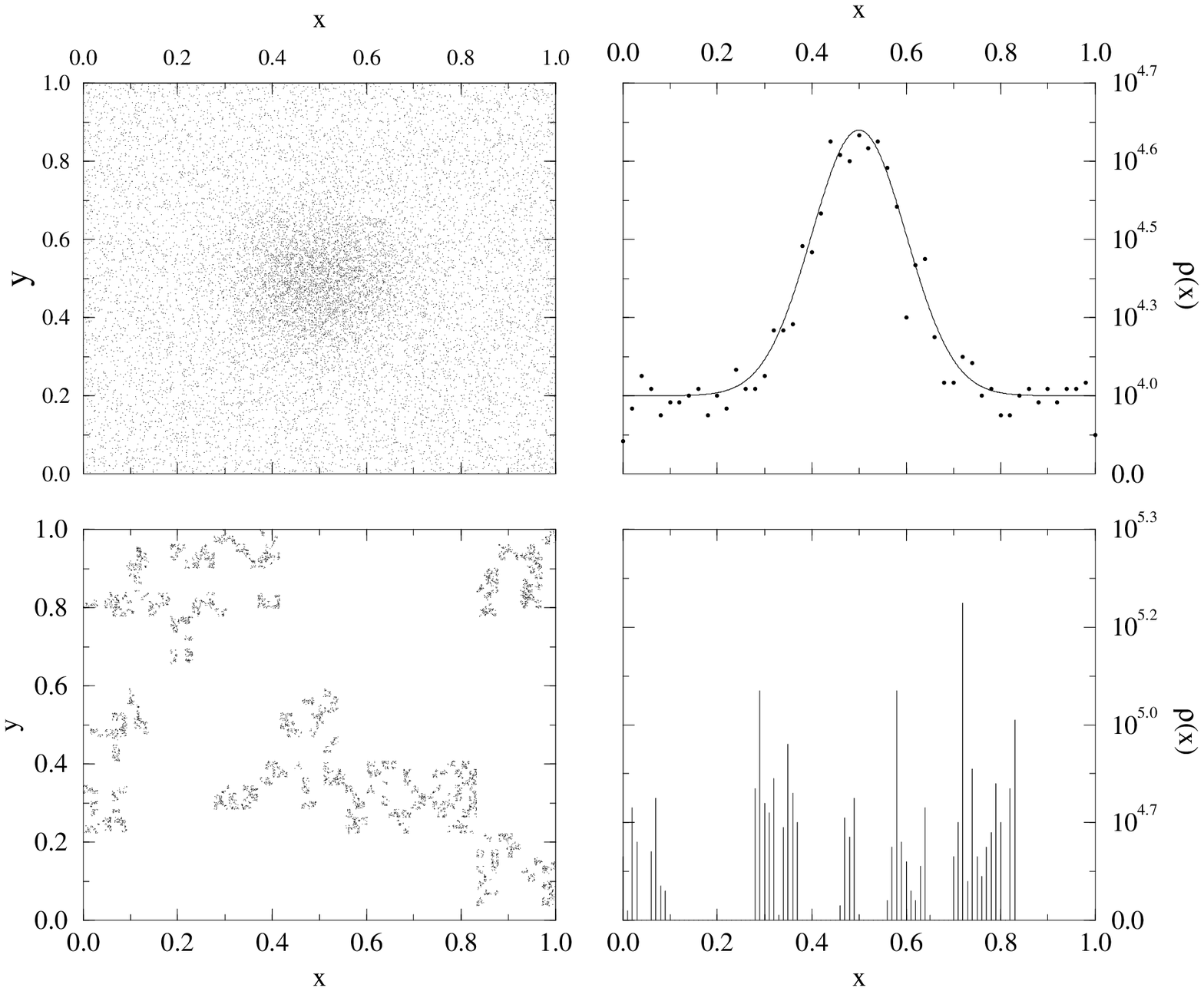}}  
\caption{\label{fig4} 
Example of analytical and nonanalytic structures. {\it Top panels}
(Left)  A cluster in a homogenous distribution. (Right)
Density profile. In this case the fluctuation
 corresponds to an enhancement
of a factor 3 with respect to the average density.
{\it Bottom panels} (Left) Fractal distribution 
in the two dimensional Euclidean space. (Right) Density profile. In this 
case the fluctuations are non-analytical and there is no 
  reference value, i.e. the average density. The average density
scales as a power law from any occupied point of the structure.
}
\eef

Most of theoretical physics is based on analytical functions and 
differential equations. This implies that structures should be 
essentially smooth and irregularities are treated as single fluctuations 
or isolated singularities. The study of critical phenomena and the 
development of the Renormalization Group (RG) theory in the 
seventies was therefore a major breakthrough \cite{wil82} \cite{amit86}.
One could observe and describe phenomena in which {\it intrinsic 
self-similar irregularities develop at all scales} and fluctuations cannot 
be described in terms of analytical functions. The theoretical methods 
to describe this situation could not be based on ordinary differential 
equations because self-similarity implies the absence of analyticity 
and the familiar mathematical physics becomes inapplicable. In some 
sense the RG corresponds to the search of a space in which the 
problem becomes again analytical. This is the space of scale 
transformations but not the real space in which fluctuations are 
extremely irregular. For a while this peculiar situation seemed to be 
restricted to the specific critical point corresponding to the competition 
between order and disorder. In the past years instead, the 
development of Fractal Geometry \cite{man83},
has allowed us to 
realize that a large variety of structures in nature are intrinsically 
irregular and self-similar (Fig.\ref{fig4}). 

Mathematically this situation corresponds to the fact that these 
structures are singular in every point.  This property can be now 
characterized in a quantitative mathematical way by the fractal 
dimension and other suitable concepts. However, given these subtle 
properties, it is clear that making a theory for the physical origin of 
these structures is going to be a rather challenging task. This is 
actually the objective of the present activity in the field 
\cite{evp95}.
The main difference between the popular fractals like coastlines, 
mountains, trees, clouds, lightnings etc. and the self-similarity of 
critical phenomena is that criticality at phase transitions occurs only 
with an extremely accurate fine tuning of the critical parameters 
involved. In the more familiar structures observed in nature, instead, 
the fractal properties are self-organized,  i.e. they develop spontaneously 
from the dynamical process. It is probably in view of this important 
difference that the two fields of critical phenomena and Fractal 
Geometry have proceeded somewhat independently, at least at the 
beginning.

The fact that we are traditionally accustomed to think in terms of 
analytical structures has a crucial effect of the type of questions we 
ask and on the methods we use to answer them. If one has never been 
exposed to the subtileness on nonanalytic structures, it is natural that 
analyticity is not even questioned. It is only after the above 
developments that we could realize that the property of analyticity 
can be tested experimentally and that it may or may not be present  
in a given physical system.

These results have important consequences from a 
theoretical point of view. In fact, when one deals 
with self-similar structures the relevant  physical 
phenomenon that leads to the scale-invariant 
structures is characterized by the {\it exponent} 
and {\it not the amplitude} of the  physical 
quantities that characterizes such distributions. 

Indeed, the only relevant and meaningful quantity is the  exponent 
of the power law correlation function (or of the space density), 
while the amplitude of the correlation  function, or of the space 
density and of the luminosity functyion \cite{slp96}, is just  related to the sample size and to 
the lower cut-offs of the distribution.  The geometric 
self-similarity has deep implications for the  non-analyticity 
of these structures. In fact, analyticity or regularity would 
imply that at some small scale,  the profile becomes smooth 
and one can define  a unique tangent. Clearly this is impossible 
in a self-similar structure because, at any small scale, a new 
structure appears and the  distribution is never smooth.
 Self-similar structures are therefore intrinsically irregular 
at all scales and correspondingly one  has to change the 
theoretical framework into one which is  capable of 
dealing with non-analytical fluctuations. This means
 going from differential equations to something like 
the  Renormalization Group to study the exponents.
 For example the so-called "Biased theory of galaxy
 formation" \cite{kai84}
is implemented considering 
the evolution of  density fluctuations within an analytic 
Gaussian framework,  while the non-analyticity of fractal 
fluctuations  implies a breakdown of the central limit 
theorem which is the  cornerstone of Gaussian processes 
\cite{pie87} \cite{cp92} \cite{evp95} \cite{bslmp94}.

 The highly  irregular galaxy distributions with large structures and 
voids strongly point to a new statistical approach in which the 
existence of a well defined average density is not assumed a priori and 
the possibility of non analytical properties should be addressed 
specifically. The new approach for the study
 of galaxy correlations in all the available catalogues 
shows that their properties are actually compatible with each other 
and they are statistically valid samples. The severe discrepancies 
between different catalogues that have led various authors to consider 
these catalogues as {\it not fair}, were due to the inappropriate methods of 
analysis.
 
The correct two point correlation analysis shows well defined fractal 
correlations up to the present observational limits, from 1 to 
$1000\hmp$ with fractal dimension $D \simeq 2$.
Of course the statistical quality and 
solidity of the results is stronger up to 
$100 \div 200 \hmp$ and 
weaker for larger 
scales due to the limited data. It is remarkable, 
however, that at these larger scales one observes exactly the continuation
of the correlation properties of the small and intermediate scales.
These results are currently at the center of acute debates in the field
and we refer to \cite{pmsl97} and \cite{dav97} for a review
of the two different points of view on this subject.

 From the theoretical point of view the fact that 
we have a situation characterized by {\it self-similar structures},  
 implies that we should not use concept
 like $\xi(r)$, $r_0$, $\delta N/N$ and certain properties of
the power spectrum, because they are not suitable to represent 
the real properties of the observed structures. 
In this respect also the N-body simulations
should be considered from a new perspective.
One cannot talk about "small" or "large" amplitudes 
for a self-similar structure because of the lack of a reference value like the 
average density.
The Physics should shift from {\it "amplitudes"} towards {\it "exponent" }
and the methods of modern statistical Physics should be adopted.
This requires the development of constructive interactions between two fields.

As we have already mentioned, the correct reanalysis of the experimental data 
is just the first step in the understanding of the
properties of the galaxy large scale structures. 
We refer the reader to Sylos Labini \etal \cite{slmp97} for a 
review of the theoretical problem. It is worth to mention that 
Sanchez \etal \cite{sanchez1,sanchez2} have proposed a 
field theory approach to the fractal structure of the universe.
In such a model the dominant dynamical mechanism responsible
for the scale invariant distribution is self-gravity itself.
Although there are several open problems, as for example the assumption of
quasi-isothermal equilibrium of galaxy distribution, 
this model represents an interesting approach and a first attempt
to focus the theoretical investigation on the behaviour of the 
exponents rather than on the amplitudes of correlations.

\section*{Acknowledgemnts}
We would like to warmly thank Prof. N. Sanchez and Prof. H. de Vega for
useful discussions and for their 
kind hospitality.

\end{document}